# Ferroelectricity in confined small particles.

E. V. Charnaya[1,2]*, A. L. Pirozerskii[2], Cheng Tien[1], and M. K. Lee[1]

[1]Department of Physics, National Cheng Kung University, Tainan 71001 Taiwan

[2]Institute of Physics, St. Petersburg State University, St. Petersburg 198504 Russia

The influence of the pore network geometry and interparticle long-distance electric coupling on the ferroelectric phase transition in small particles embedded into mesoporous matrices is considered. It was shown that the temperature of the ferroelectric phase transition in a system of electrically linked particles can be well different from that in isolated small particles. In particular, the model suggests an explanation for the weakening of size-effects on the ferroelectric phase transition in confined geometry.

*Corresponding author. E-mail: charnaya@mail.ncku.edu.tw

## 1. Introduction

Recently, a great deal of attention was focused on fabrication and studies of physical properties of various materials embedded into mesoporous matrices. In particular, ferroelectricity in confined geometry is of great interest for fundamental and applied physics. The advantages of such nanocomposites consist in screening the confined particles from harmful environment and creating particle arrays of known and well defined geometry. Experimental studies were carried out for porous glasses and alumina, artificial opals, and molecular sieves filled with sodium nitrite, Rochelle salt, TGS, and other ferroelectrics (see [1-4] and references therein). The most results obtained have shown that temperatures of the phase transitions to the ferroelectric state remained rather similar to their bulk values or even increase compared to bulk. This does not agree with predictions of theoretical models developed for isolated small ferroelectric particles (see, for instance, [5,6] and references therein). According to those models, when the particle size reduces, the temperature of the ferroelectric phase transition can decrease or increase depending on the sign of the extrapolation length $\delta$ which is specified by the particle surroundings. For small particles in vacuum or in dielectric environment the extrapolation length $\delta$ is positive. In this case the phase transition temperature decreases with decreasing the particle size and becomes zero at some critical value. The pore sizes which limit the dimensions of particles confined within the porous matrices used in experimental studies [1-4] were near or less than the critical isolated particle size. Therefore, one should expect noticeable decreasing the phase transition temperature in contrast with the experimental data. Physical reasons caused the weakening of the size-effects on the phase transitions in confined ferroelectrics were not discussed in literature. No noticeable shift of the ferroelectric phase transition was found either for dense barium titanate nanoceramics [7], while the strong reduction of the ferroelectric transition temperature was reported upon decreasing sizes of isolated BaTiO$_3$ particles (see, for instance, [8,9]). The discrepancy above mentioned raises a question whether ferroelectricity in confined particles can be examined only on the basis of size-effects in isolated small particles or some additional considerations should be taken into account. Since ferroelectricity is associated with the long-distance electric coupling, one can suggest that properties of confined ferroelectric particles should be strongly influenced by interaction between particles within neighbor pores. Here we present a theoretical model of the ferroelectric phase transition in an array of small separate particles and analyze the role of the interparticle distance and array geometry in variations of the phase transition temperature. We will show that this model can be used to treat the properties of confined ferroelectric materials.

## 2. Theoretical model

Let us consider a regular macroscopic array of small ferroelectric particles at a certain distance from each other. The Gibbs free energy of the nanoparticle system can be written as a sum of the energy of particular particles and of their coupling. To describe the phase transition in a small isolated particle an approach based on the Landau expansion of the free energy is usually used (see [5,6] and references therein), polarization playing the role of the order parameter. Since the order parameter is no longer homogeneous because of the small particle size and elevated surface-to-volume ratio, the conventional free energy expansion in terms of powers of the order parameter is supplemented with two additional terms accounting for its spatial variations and the surface contribution and with relevant boundary conditions (see, for instance, [5,6,10]). The term corresponding to the order parameter spatial variations coincides with the gradient term which is included in the Landau expansion when the order parameter fluctuations are considered [11]. Therefore, the free energy of the $i$-th small ferroelectric particle can be written as a sum of volume and surface integrals [5,6]:

$$F_i = \int \left( \frac{1}{2} A_0 (T - T_0^0) P_i^2 + \frac{1}{4} B P_i^4 + \frac{1}{6} C P_i^6 + \frac{1}{2} D (\nabla P_i)^2 \right) dV_i + \int \frac{1}{2} D \frac{P_i^2}{\delta} dS_i \qquad (1)$$

where $\vec{P}_i$ is the polarization which depends on coordinates, T is temperature, $A_0$ and $D$ are positive coefficients in the Landau expansion, B is positive or negative for the second or first order phase transition, respectively, C is equal to zero for the second order phase transition and positive for the first order one, $T_0^0$ is the phase transition temperature in the case of the second order phase transition and the lower limit of the non-polarized state for the first order phase transition, δ is the extrapollation length which determines the polarization gradient on the particle boundary according to the following law:

$$\frac{\partial \vec{P}}{\partial \vec{n}} = -\frac{\vec{P}}{\delta}, \qquad (2)$$

$\vec{n}$ is the unit normal to the particle surface. In Eq. (1) the depolarizing fields are not taken into account as also was done in the published models of the ferroelectric phase transition in an isolated particle (see [5,6] and references therein). The neglect of the depolarizing effects for isolated particles was justified by the assumption that the polarization charges on the particle surface are completely compensated with charges from outside. For ferroelectric particles embedded into porous matrices one also can suggest that the bound charges on their surface are compensated with outer charges, for instance, from the pore inner surface or from absorbed water.

To take into account the interparticle coupling we assume that the distances between ferroelectric particles exceed several unit cells. Thus, the interaction between particles is reduced to the dipole-dipole coupling [12]. The coupling energy of the *i*-th and *j*-th particles then is given by

$$F_{ij} = \int dV_i \int dV_j \left[ \frac{(\vec{P}_i \vec{P}_j)}{\rho_{ij}^3} - \frac{3(\vec{\rho}_{ij}\vec{P}_i)(\vec{\rho}_{ij}\vec{P}_j)}{\rho_{ij}^5} \right] g_i g_j, \qquad (3)$$

where $\vec{\rho}_{ij}$ is the radius-vector connected points belonging to the *i*-th and *j*-th particles, the numerical coefficients $g_i$, $g_j$ account for the decrease of the dipole coupling by free charges which compensate the polarization charges on the particle surfaces [12].

To simplify the further calculations one can use the results of Refs. [5,6] and to replace the inhomogeneous polarization in the real *i*-th particle with the effective homogeneous polarization $\vec{p}_i$ resulted from averaging over the particle volume. Then Eq. (1) transforms to

$$F_i = (\frac{1}{2}\alpha_i p_i^2 + \frac{1}{4}\beta_i p_i^4 + \frac{1}{6}\gamma_i p_i^6)V_0^i, \qquad (4)$$

where $V_0^i$ is the volume of the *i*-th particle, the coefficient $\alpha_i$ is linearly dependent on temperature $\alpha_i=\alpha_{0i}[T-T_0^i(V_0^i)]$, $\alpha_{0i}$ are positive and temperature independent coefficients, $\beta_i>0$, $\gamma_i=0$ and $\beta_I<0$, $\gamma_I>0$ for the second and first order phase transitions, respectively, $T_0^i(V_0^i)$ is the temperature of the ferroelectric phase transition in the *i*-th isolated particle, which depends on the particle shape and dimensions, for the second order phase transition or the lower limit of the non-polarized state for the first order one. The deviation of $T_0^i(V_0^i)$ from the relevant values in bulk increases with decreasing the particle size. The sign of deviation depends on the sign of the extrapolation length $\delta$. For small particles in vacuum or in dielectric environment the extrapolation length $\delta$ is positive. In this case $T_0^i(V_0^i)$ decreases with decreasing the particle size. Therefore, the temperature of the second as well as of the first order phase transition becomes zero at some critical particle size.

Within the framework of the effective polarization approximation, the dipole-dipole coupling (3) between the *i*-th and *j*-th particles can be reduced to the coupling of two point dipoles at a distance $\vec{r}_{ij}$ from each other

$$F_{ij} = [\frac{(\vec{p}_i \vec{p}_j)}{r_{ij}^3} - \frac{3(\vec{r}_{ij}\vec{p}_i)(\vec{r}_{ij}\vec{p}_j)}{r_{ij}^5}]V_0^i V_0^j g_i g_j. \qquad (5)$$

Therefore, the free energy of a system of interacting small particles is given by

$$F = \sum_i F_i + \frac{1}{2}\sum_{i \neq j} F_{ij}. \qquad (6)$$

For a large regular array of identical small particles the problem reduces to the sum of equal terms which represent the free energy $F_1$ of a single particle coupling with all other particles in the network, $F_1$ being given by

$$F_1 = (\frac{1}{2}\alpha p^2 + \frac{1}{4}\beta p^4 + \frac{1}{6}\gamma p^6)V_0 + Kp^2 V_0^2 g^2. \qquad (7)$$

The dimension factor *K* is determined by the geometry of the particle network and polarization direction. Depending on the particle lay-out, the sign of *K* can be negative or positive. In Eq. (7) the indices i are omitted because of identity of all particles. Eq. (7) leads to the following change of the phase transition temperature $\widetilde{T}_C$ in the array of coupled particles compared to the temperature of the phase transition $T_C(V_0)$ in isolated small particles:

$$\widetilde{T}_C = T_C(V_0) - \frac{2KV_0 g^2}{\alpha_0} . \tag{8}$$

Eq. (8) is valid for the ferroelectric phase transitions of both second and first orders.

## 3. Discussion

Eq. (8) shows that the temperature of the ferroelectric phase transition in the system of small particles differs from that in isolated particles. The evaluation of the factor *K* which governs the change of $\widetilde{T}_C$ compared to the phase transition temperature in isolated small particles can be carried out for some particular geometries of the small particle array.

Let us consider as an example a system of small particles which form a primitive hexagonal lattice with lattice constants *a* and *c* and assume that the polarization in particles is directed along the hexagonal *c* axis. The hexagonal geometry corresponds, for instance, to some molecular sieves like MCM-41 and SBA-15. To calculate the value of *K* we should sum up the dipole-dipole coupling of a particular particle with all other particles in the system. As is known, the result of such summation depends of the shape of the sample because of the long-distance nature of the dipole coupling [12]. We have chosen the hexagonal prisms with equal numbers of unit cells along the three crystallographic axes as a volume over which we sum the dipole coupling. Then the factor *K* was numerically computed for various ratios of *a* to *c*. A dimensionless factor $k=Ka^3$ was used in the computations to facilitate the treatment of the results obtained. The dependence of *k* on *c/a* at a constant *a* is shown in Fig.1. When *c<0.955a* the factor *k* is negative. Thus, at this geometry of the three-dimensional particle array the interparticle dipole coupling leads to a reduction of the free energy per particle and to the corresponding increase of the phase transition temperature according to Eqs. (7) and (8). The resulting temperature of the ferroelectric phase transition will be determined by the interplay of the size-effects in isolated particles and the interparticle dipole coupling.

When *c>0.955a* the chosen geometry of the interacting particle array provokes the increase of the free energy per particle and the decrease of the phase transition temperature compared to isolated small particles as it follows from Fig.1 and Eqs. (7) and (8). However, this situation seems unlikely, since the uniform polarization in particles along the same direction becomes unfavorable. Even within the framework of the model of uniaxial ferroelectric particles the spontaneous polarization vectors can have parallel and opposite mutual orientations.

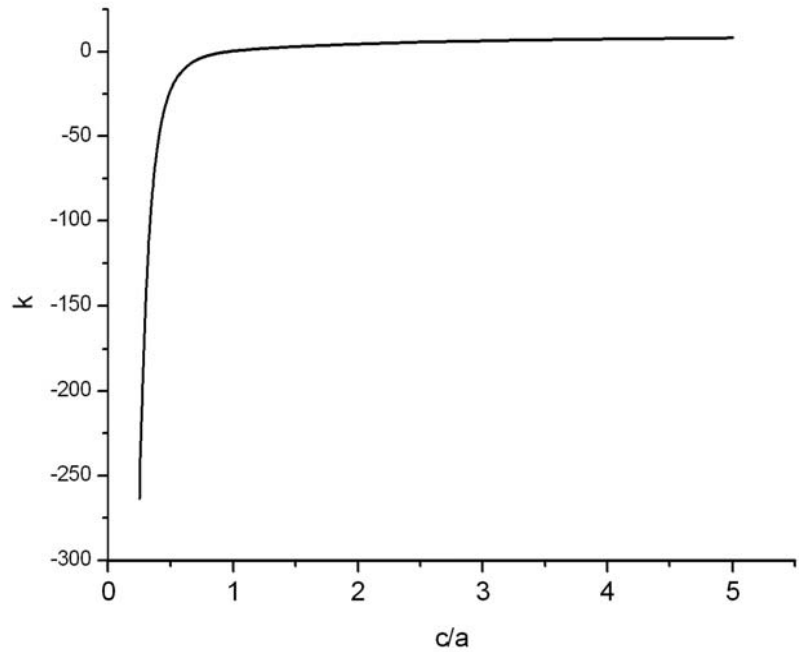

Fig.1. The dimensionless factor *k* versus c/a for the hexagonal array of small ferroelectric particles calculated at constant *a*. Circles show points calculated, the solid line is a guide for the eye.

Our computations showed that for the hexagonal array of identical particles with *c>0.955a* the interparticle interaction leads to a free energy decrease when the mutual arrangement of polarizations at particles corresponds to parallel directions along the *c* axis and alternate directions in the plane perpendicular to *c*. The similar case was discussed in [13] for a domain structure in a ferroelectric superlattice.

For large *a* and *c* tending to infinity the factor *K* tends to zero and the temperature of the phase transition approaches that in the isolated small particle. In the opposite limit of *a* and *c* tending to zero, the phase transition temperature should approach the bulk value. However, the bulk limit cannot be achieved within the framework of the point dipole approximation.

In the general case of regular or irregular particle array, which corresponds to ferroelectrics embedded into pores of opals, porous glasses and molecular sieves as well as to dense ferroelectric ceramics, the spontaneous polarization at particles can be directed arbitrarily chiefly because of arbitrary orientations of the crystallographic axes in particles. On can suggest that the interparticle coupling in this case should favor the mutual orientations of the spontaneous polarizations which yield the decrease of the particle system energy and the increase of the phase transition temperature compared to that in the isolated small particle.

The above calculations of the factor *k* for the hexagonal array of ferroelectric nanoparticles can be used to estimate the temperature of the ferroelectric phase transition in $NaNO_2$ confined within MCM-41 molecular sieves [1,4]. In particular matrices used in [1,4] the channel diameter was 3.7 nm and the channel walls were from 0.8 to 2.0 nm. Assuming the wall thickness to be 1.5 nm one can model the array of particles within pores with a hexagonal lattice with a distance between particle in the plane perpendicular to the channels *a*=5.2 nm. According to [1,4] the temperature of the ferroelectric phase transitions in confined $NaNO_2$ was near the bulk transition point. Then for small particles within pores, which diameter is near the critical particle size corresponding to the decrease of the phase transition temperature to zero, the shift $-2KV_0g^2/\alpha_0$ should be of the order of $T^0_C$ as it follows from Eq. (8). This requirement is fulfilled, for instance, when $g^2$=0.5, the ratio *c/a*=0.7, the particle size along the channels is 2.6 nm, and the pore filling is near 70 %. The Curie constant $4\pi/\alpha_0$=5000 K from [14] was used for the estimates. The fitting parameters agree well with the parameters of the filled porous matrices studied in [1,4].

In conclusion, the presented analysis of the ferroelectric phase transition in an array of electrically coupled small particles has shown that the long-distance electric coupling between small particles can increase remarkably the ferroelectric phase transition temperature compared to that in isolated small particles. The influence of the electric coupling depends on the distance between particles and geometry of the particle array. The increase in the phase transition temperature obtained for Rochelle salt confined within porous alimuna and weak size-effects on the ferroelectric phase transition observed for sodium nitrite embedded into opals and mesoporous sieves agree with the model developed above.

**Acknowledgements.** The present work was supported by Taiwan government under Grant OUA 95-21T-2-017 and by RFBR (Russia) under grants 05-02-04001 and 04-02-16159.